# Dynamic Viscosity for HFC-134a + Polyether mixtures up to 373.15 K and 140 MPa at low polyether concentration. Measurements and Modelling.

**María J.P. Comuñas[a,b*], Antoine Baylaucq[b], Christian Boned[b] and Josefa Fernández[a]**

[a]*Laboratorio de Propiedades Termofísicas, Departamento de Física Aplicada, Facultad de Física, Universidad de Santiago de Compostela, E-15782 Santiago de Compostela, Spain.*
[b]*Laboratoire des Fluides Complexes, Faculté des Sciences, BP 1155, F-64013, Pau Cedex, France.*
*\*To whom correspondence should be addressed, Laboratorio de Propiedades Termofísicas, Departamento de Fisica Aplicada, Facultad de Fisica, Universidad de Santiago de Compostela, E-15782 Santiago de Compostela, Spain, email: famajose@usc.es, Fax number 34981520676.*

**Abstract:**

This paper reports viscosity data for mixtures containing a refrigerant (HFC-134a) and a lubricant (triethylene glycol dimethyether, TriEGDME or tetraethylene glycol dimethylether, TEGDME), at 12% mass fraction of TriEGDME and 14% of TEGDME respectively. The measurements (140 data points) have been carried out versus pressure (between 10 and 140 MPa) in the monophasic liquid state from 293.15 to 373.15 K. Due to the fact that at normal pressure and temperature the HFC-134a and the polyethers are not in the same single phase (the refrigerant is a gas whereas the polyethers are liquids) accurate measurements of their mixtures require specific procedures for the samples preparation and the filling of the apparatus. A specially designed isobaric transfer falling-body viscometer is used in this work. The viscosity of the mixtures is in average 40% higher than that of the pure refrigerant, and this increase is more noticeable at low temperatures. The experimental viscosities have been used in order to check the predictive and correlation ability of several viscosity models (mixing rules, Geller and Davis method, self referencing model, hard sphere theory, free volume model and friction-theory). Most of the studied models under estimate dynamic viscosity values over all the temperature and pressure ranges.

**Introduction**

The number of works concerning the dynamic viscosity of refrigerant + lubricant mixtures has increased in the last years. Most of them involved experimental and theoretical studies on mixtures containing POEs (Polyol Esters). For example, Geller *et al.* [1-3] have reported the experimental viscosity of binary HFCs + commercial POEs mixtures in the temperature interval 253.15-393.15 K, and from the saturation pressure up to 5 MPa for lubricant mass concentration between 0.25 to 1. The same kind of systems have been studied by Jonsson and Lijle[4] with a falling ball viscometer between 313.15 and 353.15 K, up to 34 MPa, over all the composition range. The dynamic viscosity for HFCs + Pentaerythritol Esters systems has been measured by Thébault and Vamling[5] and by Wahlström and Vamling[6] with a capillary viscometer under saturation pressure. In the first paper the viscosities are reported in the temperature interval 300-350 K and over the refrigerant mass concentration from 0 to 30%, and in the later from 300 to 333 K and up to 22% of refrigerant. Cavestri and Schafer[7] have determined the viscosity for HFC-410A + commercial POE mixtures, from 258.15 to 378.15 K at the saturation pressure.

Besides, works analysing the refrigerant + PAGs (polyalkylene glycols) systems are less frequent. There are only in the literature two articles of Kumagai *et al.*[8, 9] reporting data of HFC-134a with ethyleneglycol, diethylene glycol, triethylene glycol, tetraethylene glycol and two polyethylene glycols from 273.15 to 333.15 K at the saturation pressure. It is interesting to notice that for all the above mentioned articles the maximum pressure is 34 MPa. Until now, to our knowledge, there are no experimental viscosity data in the literature of HFC-134a + polyalkylene glycol dimethylether mixtures. In the present work the first experimental dynamic viscosity values for HFC-134a + TEGDME and for HFC-134a + TriEGDME at low lubricant concentration (12% mass fraction of TriEGDME and 14% of TEGDME) are reported. The mixtures with low content in lubricant are of high interest in engineering, because in vapour compression refrigeration and heat pump systems a small amount of compressor lubricant, as a part of the working fluid, is circulating with the refrigerant[10]. The working fluid is changed from a pure refrigerant, with well defined properties, to a mixture with properties that are poorly understood and dependent on the lubricant concentration. The migrated oil affects the heat transfer coefficient, which is an important characteristic of any working fluid, changing the evaporator pressure drop and degrading the performance of both the condenser and evaporator. The knowledge of these effects is needed in the final election of the appropriate refrigerant + lubricant pair, this is the reason why these types of mixtures have been studied in several publications[11-14]. The change of the heat transfer coefficients, due to the presence of lubricant, is depending of the lubricant's type (POEs, mineral oils, PAGs or others) and of the lubricant concentration, as Sundaresan *et al.* [15] have suggested. For refrigerant + POE mixtures Geller and Davis[2] have demonstrated that their coefficients are about 50% higher than those found for refrigerant + mineral oil mixtures.

The heat transfer coefficients are related to the conduction, convection and radiation processes, which sometimes are expressed as the Nusselt or Prandtl parameters[16]. The Reynolds number is also important to determine if the flow is laminar or turbulent. In order to calculate these factors, the dynamic viscosity and the density are needed. The oil viscosity has also another important effect on the evaporator performance at low lubricant concentrations, as suggested by McMullan *et al.* [10]. Their results indicate that the degradation of the performance of the evaporator can be minimized by selecting a low viscosity oil if the oil fraction is small, whereas the lubricant should have a high viscosity if the oil concentration is high.

For all these reasons the experimental viscosity data for refrigerant + lubricant mixtures at low lubricant concentration are very interesting from an industrial point of view. The experimental viscosity data (140 points) reported in this work for HFC-134a + TriEGDME and HFC-134a + TEGDME mixtures for 12% and 14% polyether mass fraction (or 0.9295 and 0.9290 refrigerant mole fraction) respectively, and over the temperature interval of 293.15 - 373.15 K, have been extended up to high pressures (up to 140 MPa) which is more of a fundamental interest than an industrial one. The experimental viscosities have then been used in order to check the ability of different mixing rules[17, 18] and of several viscosity models (Geller and Davis method[1], self referencing model[19], hard sphere theory[20-22], free volume model[23, 24] and friction-theory[25, 26]) for viscosity prediction and correlation of refrigerant + lubricant mixtures.

**Experimental Techniques**

*Materials.* HFC-134a (molar mass 102.03 g·mol$^{-1}$) was obtained from Gazechim Froid with a purity of 99.94% and with water content not more than 24 ppm. The polyethers, TriEGDME (molar mass 178.23 g·mol$^{-1}$) and TEGDME (molar mass 222.28 g·mol$^{-1}$) were obtained from Aldrich with chemical purity better than 99%.

***Presentation of the falling-body viscometer.*** Because the lubricant and refrigerant are in two different thermodynamic states at atmospheric pressure and ambient temperature, accurate measurements of their mixtures require specific procedures for the samples preparation and the filling of the different apparatus used during this work. The mixtures are prepared in a high-pressure variable-volume cell, containing a stainless-steel ball (in order to agitate and homogenize the mixture) and equipped with a piston in order to isolate the lubricant/refrigerant mixture from the pressurizing fluid (oil). This high-pressure cell is pressurised up to a pressure higher than the saturation pressure to ensure that the mixture is in a single-phase and is homogeneous.

The dynamic viscosity has been determined with the aid of a specially designed isobaric transfer falling-ball viscometer. A detailed description of the experimental equipment has been presented in a previous work[27], in which the dynamic viscosity for methane + decane mixtures has been measured. In this apparatus, a stainless-steel cylinder falls through a fluid of unknown viscosity under selected conditions of temperature and pressure. The viscosity is a function of the falling time ($\Delta t$), of the density of both the cylinder, $\rho_c$, and the fluid, $\rho_l$, and of the apparatus parameters (a, b and c), according to the following working equation:

$$\eta = a\,(\Delta\rho\,\Delta t)^2 + b\,(\Delta\rho\,\Delta t) + c \tag{1}$$

where $\Delta\rho = \rho_c - \rho_l$. In order to determine the calibration constants in eq. 1, measurements of the falling time for reference substances, with known viscosity and density values, under similar ranges of $\Delta\rho\,\Delta t$ at the same pressure and temperature conditions are required. For each fluid the measurements of the falling time was repeated six times at thermal and mechanical equilibrium, and it is reproducible to better than 1%. The final values are the average of these measurements. The temperature is measured (with an AOIP system) with an accuracy of ± 0.5 K and the pressure (with an HBM-P3M manometer) with an accuracy within ± 0.2 MPa.

***Calibration of the falling-body viscometer.*** In this work, toluene, propane, pentane and heptane have been chosen as reference fluids in order to determine the calibration parameters (a, b, and c). The falling time has been measured for these four compounds between 293.15 to 373.15 K in steps of 20 K and from 10 to 140 MPa in 10 MPa intervals. The density and viscosity for the reference fluids, over these temperature and pressure intervals, are needed in eq. 1. Concerning the density, we have used the values reported by Vogel *et al.* [28] for propane and the Tait-like correlations reported by Cibulka *et al.* [29-31] for toluene, pentane and heptane. For viscosity, values interpolated from published experimental data of propane[28], pentane[32-34], heptane[35, 36] and toluene[22, 37] have been used. The uncertainty and the temperature and pressure intervals for the literature viscosity values, used in the calibration, are presented in Table 1. The apparatus parameters, a, b, and c are determined by plotting the reference values of $\eta$ versus $\Delta\rho\Delta t$ for each temperature. In order to verify the calibration, the dynamic viscosity of hexane has been measured for 313.15 and 333.15 K and from 10 to 140 MPa. The viscosity values obtained for hexane by using this calibration agree with the literature values published by Oliveira and Wakeham[33] within an average absolute deviation of 0.4% and a maximum deviation of 1.3%. In the following this calibration method is mentioned as method (I).

However, important discrepancies were sometimes found, between different reported viscosity measurements for the reference fluids[38]. Therefore, in order to minimize the viscosity uncertainty of our viscosity measurements, in addition to the previous method (I) two other calibration methods were used: estimated viscosity values of the reference fluids are calculated with the Hard Spheres model by Assael *et al.* [20, 22] (calibration method mentioned as HS) and also with the one-parameter friction theory model by Quiñones-Cisneros *et al.* [25, 26] (calibration method mentioned as FT). These methods have also been applied in a recent work concerning the dynamic viscosity of pure HFC-134a[38].

As a result, up to three different viscosity estimations, for each one of the reference sets of viscosities (I, HS and FT), are derived for each measured temperature and pressure. One calibration curve was made for each measured temperature and pressure and for each one of the three sets of reference viscosities. The final experimental values for viscosity are the average value of those obtained with the three methods. Taking into account that the fundamental contribution to the viscosity uncertainty is the accuracy of the calibration data, we have previously[38] estimated an accuracy of ± 2% for the 293.15, 313.15 and 333.15 K isotherms, ± 3% for the 353.15 K isotherm and ± 4% for the 373.15 K.

The viscosity of hexane has also been determined, by using for viscometer calibration these last two methods, HS and FT. The deviations obtained between the experimental data[33] and the values determined using the three calibrations methods, are plotted against the pressure in Figure 1. The experimental values obtained for hexane using the FT method are slightly higher than those determined with the HS method, but the average absolute deviation and the maximum deviation between both data sets are only 1.2% and 2%, respectively. Our viscosity values for hexane with the HS method, agree with the experimental values reported by Oliveira and Wakeham[33] within an average deviation of 2%, a Bias of –1% and a maximum deviation of 4.2%. Whereas with the FT method these deviations are 2%, -0.2% and 5%, respectively. In both cases |Bias|<AAD, i.e. there are some values which are higher than those reported by Oliveira and Wakeham, and others which are lower.

In order to determine the dynamic viscosities of HFC-134a + polyether mixtures the three calibration methods have been used and, as already mentioned, the final values were the average of the three viscosity values. For this purpose the hexane has been included as reference fluid in addition to propane, pentane, heptane, and toluene.

***Density measurements.*** In order to determine the viscosity in eq. 1 it is also necessary to know the density of the mixtures. The principle of measurement, the apparatus (Anton Paar DMA60/512P vibrating tube densimeter) and the experimental procedure for the density determination are described in details in our previous work concerning the pVT data of HFC-134a + TriEGDME[39] and HFC-134a + TEGDME[40] mixtures between 293.15 K and 373.15 K and from 10 MPa to 60 MPa.

**Experimental results**

Measurements of the dynamic viscosity, $\eta$, of HFC-134a + polyether mixtures have been carried out versus pressure (between 10 and 140 MPa in 10 MPa intervals) in the monophasic liquid state from 293.15 to 373.15 K in steps of 20 K. In order to determine the viscosities, the densities for these binary systems are needed in the $\Delta\rho$ term of eq. 1. The experimental densities for HFC-134a + TriEGDME and for HFC-134a + TEGDME mixtures have already been presented in previous works[39, 40] from 10 to 60 MPa between 293.15 and 373.15 K with an uncertainty lower than $2 \cdot 10^{-4}$ g·cm$^{-3}$. At pressures higher than 60 MPa and up to 140 MPa, the density values have been extrapolated with the procedure previously described by Et-Tahir *et al.*[41] using a Tait-like equation for the representation of density versus pressure.

The density values have been compared with the one generated by using the Patel-Teja equation of state with a quadratic mixing rule. The experimental data (up to 60 MPa) and the predicted values with the Patel-Teja equation of state (EoS) are reported in Table 2. The predicted densities up to 60 MPa are in good agreement with the experimental density values for HFC-134a + TriEGDME and HFC-134a + TEGDME. Specifically, average absolute deviations of 2.5% and 2.1% were found, in the entire composition range, between the experimental and the predicted densities for the systems containing TriEGDME and TEGDME respectively. The Tait-extrapolated data from the experimental densities and the predicted values with the Patel-Teja EoS from 60 to 140 MPa are also reported in Table 2. It can be noticed that always $\rho_{PT}$ > $\rho_{lab}$, ($\rho_{lab}$ is both the experimental data and the Tait-extrapolated ones). At pressures higher than 60 MPa the average absolute deviations between both, the Tait-extrapolated and the predicted values, are 1.9% for the binary mixture with TriEGDME and 1.5% for that containing TEGDME. The differences found between the different ways used in density determination, are not very important in viscosity determination, as an error of 1% on the density of the fluid leads[27] to a relative error of 0.2% on the viscosity, i.e. lower that the experimental uncertainty of the falling body viscometer, used in this study. In this work we have used the Tait-extrapolated values from the experimental densities but it is interesting to notice that it was also possible to use the density values predicted with an EoS. Nevertheless, it is necessary to choose an appropriate EoS that gives good volumetric predictions. For example, if the Soave Redlich-Kwong (SRK) EoS is used for density prediction of HFC-134a +polyether mixtures, an average absolute deviation of 15% is observed.

The viscosity values obtained for HFC-134a + TriEGDME and HFC-134 + TEGDME mixtures are presented in Table 3 are plotted against the pressure over the entire temperature interval in Figure 2. It can be seen in this figure that the values obtained with the three methods are in good agreement, only some discrepancies appear in the region where no experimental data are available for the reference fluids (over the 373.15 K isotherm).

In Figure 3, the dynamic viscosity for these mixtures are compared with the viscosity values of pure TriEGDME, TEGDME and HFC-134a determined in previous works[38, 42] over the pressure interval of 20 to 100 MPa and from 293.15 to 353.15 K. In this figure the dynamic viscosity are plotted against the temperature and the pressure in a 3D diagram (Tpη) with the same scale. The refrigerant viscosity is typically 2-3 orders of magnitude smaller than that of the lubricant. The viscosity of the mixtures is in average 40% higher than that of the pure refrigerant, and this increase is more noticeable at lower temperatures.

**High Pressure Viscosity Modelling**

In order to predict the dynamic viscosities for HFC-134a + polyether mixtures we have used some models that only involve the viscosity data of the pure compounds of the mixture. For pure HFC-134a the experimental viscosity values between 293.15 and 373.15 K and from 10 to 140 MPa have been taken from a previous work[38]. For TriEGDME and TEGDME the values already measured[42] between 293.15 and 353.15 K and from 0.1 to 100 MPa in steps of 20 K have been used. Taking into account the temperature and pressure intervals for which the viscosity of pure compounds is known, in this work the ability of these models in the temperature interval of 293.15 – 353.15 K and over the pressure range of 20 to 100 MPa has been analyzed.

In order to assess and compare the performances of various models it is necessary to introduce quantities characteristic of the results obtained. We have used the Absolute Average percentual Deviation (AAD), the Maximum percentual Deviation (DMAX), and the Average percentual Deviation (Bias) which are defined as follows:

$$AAD = \frac{100}{N} \sum_{i=1}^{n} \left| \frac{\phi_i^{exp} - \phi_i^{cal}}{\phi_i^{exp}} \right|$$

$$DMAX = Max\left( 100 \left| \frac{\phi_i^{exp} - \phi_i^{cal}}{\phi_i^{exp}} \right| \right)$$

$$Bias = \frac{100}{N} \sum_{i=1}^{n} \frac{\phi_i^{exp} - \phi_i^{cal}}{\phi_i^{exp}}$$

where $\phi_i^{exp}$ and $\phi_i^{cal}$ are, respectively, the experimental and the calculated values, and n is the total number of points.

*Mixing Rules.* In order to model the properties of the mixtures it is essential to consider the use of several mixing rules. There are different mixing rules but in this work only some simple ones are tested in order to analyse the possibility to estimate the viscosity of HFC-134a + polyether mixtures with the knowledge only of the viscosities of pure substances and their mole fraction (or composition). We will then focus only on three mixing rules without any adjustable parameter so they can be considered as predictive, and on a mixing rule that involves one adjustable parameter, i.e. a correlation mixing rule.

The first mixing rule used has the following form for a n-compounds mixture:

$$\ln \nu_m = \sum_{i=1}^{n} x_i \ln \nu_i \tag{2}$$

where $\nu_m$ and $\nu_i$ are respectively the kinematic viscosities of mixture and of pure compounds, and $x_i$ is the mole fraction of compound i. Geller and Davis[1] have utilized this mixing rule in order to predict the viscosity for refrigerant + POE lubricant mixtures. These authors have found that this method is more appropriate for high temperatures. Thus, the lower the temperature, the greater the deviations between experimental data and calculated values using eq. 2.

For HFC-134a + TriEGDME mixtures (with 12% mass fraction or 0.0705 mole fraction of TriEGDME) we have obtained an AAD of 19.5 %, a Bias of 19.5% and a DMAX of 24.4% between the calculated values of dynamic viscosity and the experimental ones of Table 3. These deviations are respectively 18.4%, 18.4% and 22.3% for HFC-134a + TEGDME system (with 14% mass fraction or 0.0710 mole fraction of TEGDME). With this mixing rule the calculated values are always lower than the experimental data (AAD=Bias).

The mixing rule proposed by Katti-Chaudhri[17] has also been used for viscosity prediction. This mixing rule has the following form for a n-compounds mixture:

$$\ln \nu_m \eta_m = \sum_{i=1}^{n} x_i \ln \nu_i \eta_i \tag{3}$$

where $\nu_m$ and $\nu_i$ are respectively the molar volumes of the mixture and of the pure compounds. The dynamic viscosity values calculated with this mixing rule have been compared with the experimental data of Table 3 for HFC-134a + TriEGDME mixture and an AAD of 20.4%, a Bias of 20.4% and a DMAX of 25.3% have been obtained.

These deviations are respectively 20.4%, 20.4% and 24.2% for HFC-134a + TEGDME mixture.

Finally we have used a mixing rule proposed by Grunberg-Nissan[18] that includes one adjustable parameter, $d_{12}$, that reflects the interactive effects. This mixing rule has the following expression for a binary mixture:

$$\ln \eta_m = \sum_{i=1}^{2} x_i \ln \eta_i + x_1 x_2 d_{12} \qquad (4)$$

If we analyse the predictive capability of this mixing rule by taking $d_{12}=0$ an AAD of 18.5%, a Bias of 18.5% and a DMAX of 22.9% are found for HFC-134a + TriEGDME mixture and for the system containing TEGDME these values are 17.0%, 17.0% and 20.2% respectively. With $d_{12} \neq 0$ the obtained results with this mixing rule are better but in this case, the method is not predictive any longer. By minimizing the average absolute deviation between the calculated and the experimental data of Table 3 for HFC-134a + polyether mixtures, we have obtained the following values for the adjustable parameter: $d_{12}=1.0547$ for the mixture that contains TriEGDME and $d_{12}=1.0129$ for the HFC-134a + TEGDME system. For the first mixture, the AAD is 6.3%, the Bias 2.3% and DMAX 14.4%. For the second one, these values are, respectively, 4.6%, -1.7% and 12.1%.

*Geller and Davis[1] prediction method.* These authors have applied a simple method for viscosity prediction in which the viscosity is related to the molar volume of the fluid by

$$\frac{1}{\eta} = \frac{B(V - V_0)}{V_0} \qquad (5)$$

where V is the molar volume, B a coefficient and $V_0$ is a hypothetical molar volume corresponding to the disappearance of the molecular transport ($\lim 1/\eta \rightarrow 0$). One of the most successful applications of this model to liquid mixtures was described by Liu and Wang[43], who obtained an average absolute deviation less than 4% for 60 binary mixtures containing polar and nonpolar compounds. Once calculated B and $V_0$ parameters for pure compounds these authors have used the following mixing rules:

$$V_{0m} = \sum_{i=1}^{n} x_i V_{0i} \qquad (6)$$

$$\ln B_m = \sum_{i=1}^{n} q_i \ln B_i \qquad (7)$$

where the subscripts m and i denote quantities for the mixture and the pure components respectively, $x_i$ is the mole fraction of component i, and $q_i = x_i V_i / V_m$. Geller and Davis[1] have utilized this model for HFC-134a + commercial POE mixtures obtaining a root mean square deviation of 15% and a maximum deviation of 30% between the calculated values and the experimental data. The HFC-32 + commercial POE systems have been studied by the same authors[2] obtaining in this case an AAD of 7% and a DMAX of 20%. This method has been used by Geller and Davis[1] to calculate viscosity of refrigerant + POE mixtures at small oil concentrations in order to estimate the effect of different lubricants on the heat transfer coefficients during the evaporation.

The ability of this method has been analysed in this work for viscosity prediction of HFC-134a + polyether mixtures. Taking into account the viscosity values previously reported for pure HFC-134a[38] and for pure TriEGDME and TEGDME[42], the values of B and $V_0$ have been determined by minimizing the average absolute deviation between the calculated and the experimental data of pure compounds. The parameter values are reported in Table 4 with the results obtained for the pure compounds. The obtained deviations between the predicted values, obtained by using eqs. 6-7, and the experimental data were an AAD of 18.5%, and Bias of 9.4% and a DMAX of 48.7% for the system containing TriEGDME. These values are 15.7%, 1.1% and 41.5% respectively for the mixture with TEGDME. The AAD obtained in this work is of the same order of magnitude than those found by Geller and Davis for HFC-134a + POE lubricants[1], but the DMAX obtained in this work is higher. However, it is necessary to point out that whereas Geller and Davis goes up to 5 MPa, in this work the predicted values have been calculated up to 100 MPa.

***The self-referencing method.*** This model has been developed by Kanti *et al.*[19], and has the advantage of only requiring one experimental value $\eta(p_0,T_0)$, at a pressure $p_0$, and a temperature $T_0$. This is the reason why this method is referred to as self-referencing model. The method involves neither molar mass nor any other physical properties (including critical parameters), more details can be found in the original paper[19]. It can similarly be applied without restriction to pure substances, synthetic mixtures and even chemically very rich systems such as petroleum cuts for which the method was originally developed. The method involves nine parameters (a, b, …, i) originally determined by Kanti *et al.*[19] using numerical analysis on a database containing linear alkanes and alkylbenzenes. On the basis of knowledge of the values of these coefficients, the method can be used directly without further adjustment, and for this reason, it may be considered as general and predictive. The formulation of this method is as follows:

$$\ln\left(\frac{\eta(p,T)}{\eta(p_0,T_0)}\right) = (ay^2 + by + c)\ln\left(1 + \frac{(p-p_0)}{dy^2 + ey + f}\right) + (gy_0^2 + hy_0 + i)\left(\frac{1}{T} - \frac{1}{T_0}\right) \quad (8)$$

where $y = y_0 + (gy_0^2 + hy_0 + i)(1/T - 1/T_0)$ and $y_0 = \ln[\eta(0.1,T_0)]$. This equation is used with p in MPa, $T_0$ and T in K and $\eta(p_0,T_0)$ in mPa·s.

In the present work we have used this formulation in order to determine the viscosity values for HFC-134a + polyether mixtures. For this objective, in a first step the parameters (a, b, ..., i) determined in the original paper by Kanti *et al.*[19] have been used in eq. 8. Different combinations have been considered for the reference temperature and pressure. In these conditions the DMAX obtained with the 168 experimental points of Table 3 (170 from which we subtract one reference point for each mixture) was important, reaching the value of 51%. This fact shows, as it could have been expected, that the original parameters are not appropriate for refrigerant + lubricant mixtures. Secondly, the following mixing rules have been used in order to calculate the mixture parameters ($a_m$, $b_m$, ..., $i_m$) from the parameters estimated on the pure compounds (HFC-134a, TriEGDME and TEGDME) viscosity data:

$$\alpha_m = \sum_{i=1}^{n} x_i \alpha_i \qquad (9)$$

where $\alpha$ represent each one of the parameters (a, b,..., i), and $x_i$ is the mole fraction of compound i. We have chosen 293.15 K and 20 MPa, respectively, as reference temperature and pressure. For polyether molecules, the parameters values have been published in previous work[44]. In table 5 we report the values of the parameters for pure HFC-134a the parameters have been fitted using our experimental viscosity values[38] and minimizing the AAD between the experimental and the calculated data. We report also in Table 5 the deviation results obtained for this pure compound. With these parameters the self-referencing model correlates the viscosity values of pure HFC-134a with an AAD of 0.5%, a Bias of 0.01% and a DMAX of 1.0%. With this method the predicted viscosities for HFC-134a + TriEGDME agree with the experimental data of Table 3, with an AAD of 4.1%, a Bias of –0.2% and a DMAX of 12.6%. These values are 5.2%, -3.2%, and 21.9%, respectively, for the mixture that contains TEGDME. The obtained AAD are quite lower than that obtained with the other methods presented in this article.

***The hard-sphere model.*** This model has been developed[45, 46] for the simultaneous correlation of self-diffusion, viscosity and thermal conductivity of dense fluids. The transport coefficients of real dense fluids expressed in terms of $V_r = V/V_0$ with $V_0$ the close-packed volume and V the molar volume, are assumed to be directly proportional to the values given by the exact hard-sphere theory. The proportionality factor, described as a roughness factor $R_\eta$, accounts for molecular roughness and departure from molecular sphericity. Universal curves for the viscosity were developed and expressed as:

$$\ln\left(\frac{\eta^*_{exp}}{R_\eta}\right) = \sum_{i=0}^{7} a_{\eta i}(1/V_r)^i \quad \text{with} \quad \eta^*_{exp} = 6.035 \cdot 10^8 \left(\frac{1}{MRT}\right)^{1/2} \eta_{exp} V^{2/3} \quad (10)$$

The coefficients $a_{\eta i}$ are universal, independent of the chemical nature of the compound, and $V_0$ and $R_\eta$ are adjustable parameters. Assael *et al.* give correlation formulas relative to $V_0$ and $R_\eta$ for alkanes[20] and aromatics[22]. $R_\eta$ is independent of pressure and temperature, while $V_0$ depends on temperature. This method can be applied to mixtures, knowing $V_0$ and $R_\eta$ parameters for each compound, and using the following mixing rules in order to determine these parameters for mixtures:

$$V_{0m}(T,x) = \sum_{i=1}^{n} x_i V_{0i}(T) \quad (11)$$

$$R_{\eta m} = \sum_{i=1}^{n} x_i R_{\eta i} \tag{12}$$

where i and m subscript are used for pure compounds and mixture, respectively.

In this work we have used the $a_{\eta i}$ universal parameters reported by Assael *et al.* [20]. The values of $V_0$ and $R_\eta$ for the two pure polyethers (TriEGDME and TEGDME) have been published in a previous paper[44]. In a recently paper Assael *et al.* [47] recommend a value of $R_\eta=1$ for pure HFC-134a, taking into account this fact the $V_0(T)$ values have been determined by minimizing the average absolute deviation between the experimental[38] and the calculated viscosities. The following expression has been obtained:

$$V_0(T) = 2.831\,10^{-5} + 6.9383\,10^{-6}\,T_r + 8.6750\,10^{-6}/T_r \tag{13}$$

where $T_r$ is the reduced temperature. The hard-sphere scheme represents the viscosity of HFC-134a with an AAD of 2.9%, with a Bias of 0.2% and with a DMAX of 7.4%.

Taking into account the parameters values for pure TriEGDME, TEGDME and HFC-134a, and considering the mixing rules eqs. 11-12, the dynamic viscosity of HFC-134 + TriEGDME and HFC-134 + TEGDME can be predicted. The temperature dependence of $V_0$ parameters of mixtures and pure compounds can be observed in Figure 4. The hard-sphere model predicts the dynamic viscosity of HFC-134a + TriEGDME mixtures with an AAD of 12.3%, a Bias of 12.3% and a DMAX of 18.9%. These values are, respectively, 5.0%, 4.0%, and 10.5% for the system containing TEGDME. Finally, if $V_0$ and $R_\eta$ are fitted for each mixture it is obtained an AAD of 1% and a DMAX of 3.1% for HFC-134a + TriEGDME and 1% and 8.5% respectively for HFC-134a + TEGDME, but in this case the model is not predictive.

***The free-volume viscosity model.*** Recently an approach in order to model the viscosity of Newtonian fluids (in the condensed phase; density ρ > 200 kg.m$^{-3}$) with small molecules has been proposed by Allal *et al.* [23]. This approach connects viscosity, η, to molecular structure via a representation of the free volume fraction. In its first version the model[23] could be only applied to dense fluids but a version valid for low density states has also been developed[24]. In this last version the viscosity has the following expression:

$$\eta = \eta_0 + \frac{\rho \ell \left( \alpha \rho + \frac{pM}{\rho} \right)}{\sqrt{3RTM}} \exp\left[ B \left( \frac{\alpha \rho + \frac{pM}{\rho}}{RT} \right)^{3/2} \right] \tag{14}$$

where M is the molar mass, ρ is the density, $\eta_0$ is the diluted gas viscosity term, for which we have used the expression proposed by Chung *et al.*[48] and $\ell$, α and B are adjustable parameters for each pure fluid. This model can be applied also for mixtures using the following mixing rules:

$$\ln \eta_{0m} = \sum_{i,j=1}^{n} x_i \ln \eta_{0i} \tag{15}$$

$$\alpha_m = \sum_{i,j=1}^{n} x_i x_j \alpha_{ij} \qquad \text{with } \alpha_{ij} = \sqrt{\alpha_i \alpha_j} \tag{16}$$

$$\ell_m = \sum_{i=1}^{n} x_i \ell_i \tag{17}$$

$$1/B_m = \sum_{i=1}^{n} x_i / B_i \tag{18}$$

where i and m subscript are used for the pure compounds and the mixture, respectively. As in above calculation, the experimental dynamic viscosity values for TEGDME,

TriEGDME and HFC-134a have been taken from previous works[38, 42]. In Table 6 the parameter values for pure compounds (TriEGDME, TEGDME and HFC-134a) are reported together with the deviations with which this model represents de dynamic viscosity of these compounds. Taking into account these parameters values and the mixing rules of eqs. 15-18 the dynamic viscosity of HFC-134a + TriEGDME and HFC-134a + TEGDME mixtures can be predicted. The calculated values with the free volume model agree with the experimental data of Table 3, within an AAD of 13.5%, a Bias of 13.5% and a DMAX 18.5% for the HFC-134a + TriEGDME mixture. These values are 15.1%, 15.1% and 20.6% for the mixture that contains TEGDME. If the $\ell_m$, $\alpha_m$ and $B_m$ values are fitted against the experimental viscosities of the mixtures, these deviations are much lower (1%, 0.3%, 6% and 0.8%, -0.2% and 2.2%, respectively) but, then, the model is not predictive.

*Friction-theory.* This model has been developed recently by Quiñones-Cisneros *et al.*[25, 26] and express the total viscosity, $\eta$, as the addition of a dilute gas term $\eta_0$ and a friction term $\eta_f$ as follows $\eta = \eta_0 + \eta_f$. The dilute viscosity is defined as the viscosity at zero density and can be obtained by the model proposed by Chung *et al.*[48]. The friction term has been linked to the van der Waals repulsive pressure term $p_r$ and attractive pressure term $p_a$ by three temperature dependent friction coefficients $\kappa_r$, $\kappa_a$ and $\kappa_{rr}$ as follows:

$$\eta = \kappa_r p_r + \kappa_a p_a + \kappa_{rr} p_r^2 \tag{19}$$

The repulsive and attractive pressure terms can be obtained from cubic equations of state (EoS); in this work the Peng-Robinson EoS has been used. In the first version of this theory[25] the viscosity of each pure compounds is modelled with the following expressions for the friction coefficients:

$$\kappa_r = a_0 + a_1(\exp[\Gamma-1]-1) + a_2(\exp[2(\Gamma-1)]-1)$$
$$\kappa_a = b_0 + b_1(\exp[\Gamma-1]-1) + b_2(\exp[2(\Gamma-1)]-1) \quad (20)$$
$$\kappa_{rr} = c_2(\exp[2\Gamma]-1)$$

with $\Gamma=T_c/T$, where $T_c$ is the critical temperature. For pure TriEGDME, TEGDME and HFC-134a the critical constants values have been published in previous works[39, 44]. In Table 7 the $a_i$, $b_i$, $c_i$ parameters and the deviations between the calculated and the experimental data[38, 42] for pure compounds are reported. Once known the parameters for pure TriEGDME, TEGDME and HFC-134a the dynamic viscosity for HFC-134a + TriEGDME and HFC-134a + TEGDME mixtures can be predicted, taking into account the following mixing rules[25, 26]:

$$\ln \eta_o = \sum_{i=1}^{n} x_i \ln \eta_{0i} \qquad \kappa_r = \sum_{i=1}^{n} z_i \kappa_{ri} \quad (21)$$

$$\kappa_a = \sum_{i=1}^{n} z_i \kappa_{ai} \qquad \kappa_{rr} = \sum_{i=1}^{n} z_i \kappa_{rri} \quad (22)$$

where:

$$z_i = \frac{x_i}{M_i^\varepsilon M_m} \qquad M_m = \sum_{i=1}^{n} \frac{x_i}{M_i^\varepsilon} \quad (23)$$

The $\varepsilon$ parameter that appears in the last equations has been fitted[26] against several compounds finding the best results with $\varepsilon=0.30$. With the friction theory we have obtained for HFC-134a + TriEGDME mixture an average deviation of 7.6%, a Bias of 7.6% and a maximum deviation of 11.5% between the predicted values and the experimental data of Table 3. These values are, respectively, 8.5%, -8.5% and 18.0% for the system containing TEGDME.

***Comparison between the models.*** The results obtained for dynamic viscosity prediction, of HFC-134 + TriEGDME and HFC-134 + TEGDME mixtures, with the different tested models: mixing rule (eq.2), Geller and Davis method, self-referencing model, hard-sphere scheme, free volume and friction theory, are compared in Figure 5. It is interesting to notice that the three models with physical background, free volume, hard-spheres and friction-theory, predict in most of cases dynamic viscosity values that are lower than the experimental values over all the temperature and pressure ranges, except for the prediction with the friction-theory for HFC-134a + TEGDME for which the predicted values are higher than the experimental data of Table 3. Furthermore, in most cases, we obtain AAD = Bias, i.e. all the points are on the same side of the experimental curve (lower values), except for the prediction with the self-referencing method and the hard-sphere scheme for HFC-134a + TEGDME mixture, where the Bias is slightly lower than the AAD, as it can be observed in Figure 5. It is interesting to point out that the three models with physical background, together with the self-referencing method, are the most adequate procedures in order to predict the viscosity of HFC-134a + polyether mixtures, for low concentration in lubricant. An extended study over all the concentration range would be useful in order to make clear the prediction ability of the representative models at higher lubricant concentration.

**Conclusion**

Dynamic experimental viscosity data for HFC-134a + TriEGDME and HFC-134 + TEGDME mixtures have been measured between 293.15 and 373.15 K and from 10 to 140 MPa, and at low polyether concentration (12% mass fraction for TriEGDME and 14% mass fraction for TEGDME) by using an specially designed isobaric transfer falling-body viscometer. The viscosity of the mixtures is in average 40% higher than that of the pure refrigerant, and this increase is more noticeable at lower temperatures. The experimental viscosities reported in this work have been used in order to check the ability of some mixing rules and of several viscosity models, (the Geller and Davis model, the self referencing model, the hard sphere scheme, the free volume model and the friction-theory), for viscosity prediction and correlation of refrigerant + lubricant mixtures. Most of the studied models under estimate dynamic viscosity values over all the temperature and pressure ranges. Other experimental measurements over wide temperature, pressure, and composition range for different refrigerant + lubricant mixtures are needed in order to complete this study.

**Acknowledgments**

This work was carried out under the Research Project PPQ2001-3022 MCYT-Spain and the Spanish-French Joint Action HF 2001-0101.

**Literature cited**

(1) Geller, V. Z.; Davis K. E., Solubility and Viscosity of Refrigerant/POE Lubricant Mixtures, presented in Int. Congr. Refrig.: Refrig. 19th, **1995**, 223.

(2) Geller, V. Z.; Paulaitis M. E.; Bivens D. B.; Yokozeki A. Viscosities of HFC-32 and HFC-32/Lubricant Mixtures. *Int. J. Thermophys.* **1996**, *17*, 75.

(3) Geller, V., Thermophysical Properties of Alternative Refrigerant/Lubricant Mixtures, presented in IIR Proceedings Series "Refrigeration Science and Technology", **1998**, 49.

(4) Jonsson, U. J.; Lilje K. C., Elastohydrodynamic Lubrication Properties of Polyol Ester Lubricants-R134a Mixtures, presented in Proceedings of 1998 International Compressor Engineering Conference at Purdue, Purdue University U.S.A., July 14-17, **1998**, 123.

(5) Thébault, C.; Vamling L., A New Correlation For Viscosity of Oil/Refrigerant Mixtures, presented in Int. Congr. Refrig.: Refrig. Third Millennium, 20th, **1999**, 707.

(6) Wahlström, A.; Vamling L., Viscosity for Mixtures of HFCs and Pentaerythritol Esters, presented in Int. Congr. Refrig.: Refrig. Third Millennium, 20th, **1999**, 716.

(7) Cavestri, R. C.; Schafer W. R. Measurement of Solubility, Viscosity, and Density of R410A Refrigerant/Lubricant Mixtures. *ASHRAE Transactions*. **2000**, *106*, 277.

(8) Kumagai, A.; Miura-Mochida H.; Takahashi S. Liquid Viscosities and Densities of HFC-134a + Glycol Mixtures. *Int. J. Thermophys.* **1993**, *14*, 45.

(9) Kumagai, A.; Miura-Mochida H.; Takahashi S. Revised Viscosities for HFC-134a + Glycol Mixtures from 273 to 333 K. *Int. J. Thermophys.* **1994**, *15*, 109.

(10) McMullan, J. T.; Hewitt N. J.; Masson A. L.; Murphy N. E. The influence of oil viscosity and refrigerant quality on evaporator performance. *Int. J. Energy Research*. **1992**, *16*, 567.

(11) Eckels, S. J.; Pate M. B. In-Tube Evaporation and Condensation of Refrigerant Lubricant Mixtures of HFC134a and CFC12. *ASHRAE Transactions*. **1991**, *97*, 62.

(12) Schlager, L. M.; Pate M. B.; Bergles A. E. Evaporation and Condensation of Refrigerant-

Oil Mixtures in Smooth Tube and Micro-Fin Tube. *ASHRAE Transactions*. **1988**, *94*, 149.

(13) Sur, B.; Azer N. Z. Effect of Oil on Heat Transfer and Pressure Drop During Condensation of Refrigerant-113 Inside Smooth and Internally Finned Tubes. *ASHRAE Transactions*. **1991**, *97*, 365.

(14) Tichy, J. A.; Macken N. A.; Duval W. M. B. An Experimental Investigation of Heat Transfer in Forced Convection Condensation of Oil-Refrigerant Mixture. *ASHRAE Transactions*. **1985**, 91,

(15) Sundaresan, S. G.; Pate M. B.; Doerr T. M., A Comparison of the Effects of Different Lubricants on the In-Tube Evaporator of an HFC-Blend Refrigerant, presented in Proceedings of 1994 International Compressor Engineering Conference at Purdue, Purdue University U.S.A., July 14-17, **1994**, 323.

(16) Lee, K. S.; Kim W. S.; Lee T. H.; Lee S. Y., An Experimental Study on the Behavior of Frost Formation in the Vertical Plate Heat Exchanger, presented in Proceedings of 1994 International Compressor Engineering Conference at Purdue, Purdue University U.S.A., July 14-17, **1994**, 329.

(17) Katti, P. K.; Chaudry M. M. Boiling points and surface tensions of mixtures of benzyl acetate with dioxane, aniline, and m-cresol. *J. Chem. Eng. Data*. **1964**, *9*, 442.

(18) Grunberg, L.; Nissan A. H. Mixture law for viscosity. *Nature*. **1949**, *164*, 799.

(19) Kanti, M.; Zhou H.; Ye S.; Boned C.; Lagourette B.; Saint-Guirons H.; Xans P.; Montel F. Viscosity of liquid hydrocarbons, mixtures and petroleum cuts, as a function of pressure and temperature. *J. Phys. Chem.* **1989**, *93*, 3860.

(20) Assael, M. J.; Dymond J. H.; Papadaki M.; Patterson P. M. Correlation and Prediction of Dense Fluid Transport Coefficients. I. Alkanes. *Int. J. Thermophys.* **1992**, *13*, 269.

(21) Assael, M. J.; Dymond J. H.; Papadaki M.; Patterson P. M. Correlation and prediction of dense fluid transport coefficients. III. n-Alkane mixtures. *Int. J. Thermophys.* **1992**, *13*,

659.

(22) Assael, M. J.; Dymond J. H.; Patterson P. M. Correlation and Prediction of Dense Fluid Transport Coefficients. V. Aromatic Hydrocarbons. *Int. J. Thermophys.* **1992**, *13*, 895.

(23) Allal, A.; Moha-Ouchane M.; Boned C. A new free volume model for dynamic viscosity and density of dense fluids versus pressure and temperature. *Phys. Chem. Liquids.* **2001**, *39*, 1.

(24) Allal, A.; Boned C.; Baylaucq A. Free-volume viscosity model for fluids in the dense and gaseous states. *Phys. Rev. E.* **2001**, *64*, 011203/1.

(25) Quiñones-Cisneros, S. E.; Zéberg-Mikkelsen C. K.; Stenby E. H. The friction theory (f-theory) for viscosity modelling. *Fluid Phase Equilib.* **2000**, *169*, 249.

(26) Quiñones-Cisneros, S. E.; Zéberg-Mikkelsen C. K.; Stenby E. H. One parameter friction theory models for viscosity. *Fluid Phase Equilib.* **2001**, *178*, 1.

(27) Daugé, P.; Baylaucq A.; Marlin L.; Boned C. Development of an Isobaric Transfer Viscometer Operating up to 140 MPa. Application to a Methane + Decane System. *J. Chem. Eng. Data.* **2001**, *46*, 823.

(28) Vogel, E.; Kuchenmeister C.; Bich E.; Laesecke A. Reference Correlation of the Viscosity of Propane. *J. Phys. Chem. Ref. Data.* **1998**, *27*, 947.

(29) Cibulka, I. Saturated liquid densities of 1-alkanols from C1 to C10 and n-alkanes from C5 to C16: a critical evaluation of experimental data. *Fluid Phase Equilib.* **1993**, *89*, 1.

(30) Cibulka, I.; Hnedkovsky L. Liquid Densities at Elevated Pressures of n-Alkanes from C5 to C16: A Critical Evaluation of Experimental Data. *J. Chem. Eng. Data.* **1996**, *41*, 657.

(31) Cibulka, I.; Takagi T. P-r-T Data of Liquids: Summarization and Evaluation. 5. Aromatic Hydrocarbons. *J. Chem. Eng. Data.* **1999**, *44*, 411.

(32) Kiran, E.; Sen Y. L. High-pressure viscosity and density of n-alkanes. *Int. J. Thermophys.* **1992**, *13*, 411.

(33) Oliveira, C. M. B. P.; Wakeham W. A. The viscosity of five liquid hydrocarbons at

pressures up to 250 MPa. *Int. J. Thermophys.* **1992**, *13*, 773.

(34) Stephan, K.; Lucas K., *Viscosity of Dense Fluids,* Plenum Press, New York and London; 1979.

(35) Kashiwagi, H.; Makita T. Viscosity of twelve hydrocarbon liquids in the temperature range 298-348 K at pressures up to 110 MPa. *Int. J. Thermophys.* **1982**, *3*, 289.

(36) Assael, M. J.; Oliveira C. P.; Papadaki M.; Wakeham W. A. Vibrating-wire viscometers for liquids at high pressures. *Int. J. Thermophys.* **1992**, *13*, 593.

(37) Santos, F. J. V. d.; Castro C. A. N. d. Viscosity of Toluene and Benzene under High Pressures. *Int. J. Thermophys.* **1997**, *18*, 367.

(38) Comuñas, M. J. P.; Baylaucq A.; Quiñones-Cisneros S. E.; Zéberg-Mikkelsen C. K.; Boned C.; Fernández J. Viscosity Measurements and Correlations for 1,1,1,2-tetrafluoroethane (HFC-134a) up to 140 MPa. *Fluid Phase Equilib.* **2003**, *210*, 21.

(39) Comuñas, M. J. P.; Fernández J.; Baylaucq A.; Canet X.; Boned C. PρTx Measurements for HFC-134a + Triethylene Glycol Dimethylether System. *Fluid Phase Equilib.* **2002**, *199*, 185.

(40) Comuñas, M. J. P.; Baylaucq A.; Boned C.; Canet X.; Fernández J. High-Pressure volumetric behaviour of x 1,1,1,2-tetrafluoroethane + (1-x) 2,5,8,11,14-pentaoxapentadecane (TEGDME) mixtures. *J. Chem. Eng. Data*. **2002**, *47*, 233.

(41) Et-Tahir, A.; Boned C.; Lagourette B.; Xans P. Determination of the viscosity of various hydrocarbons and mixtures of hydrocarbons versus temperature and pressure. *Int. J. Thermophys.* **1995**, *6*, 1309.

(42) Comuñas, M. J. P.; Baylaucq A.; Boned C.; Fernández J. High Pressure Measurements of the Viscosity and Density of Two Polyethers and Two Dialkyl Carbonates. *Int. J. Thermophys.* **2001**, *22*, 749.

(43) Liu, H.; Wang W. Application of the Hildebrand Fluidity Equation to Liquid Mixtures. *Ind. Eng. Chem. Res.* **1991**, *30*, 1617.

(44) Baylaucq, A.; Comuñas M. J. P.; Boned C.; Allal A.; Fernández J. High pressure viscosity and density modelling of two polyethers and two dialkyl carbonates. *Fluid Phase Equilib.* **2002**, *199*, 249.

(45) Assael, M. J.; Dymond J. H.; Papadaki M.; Patterson P. M. Correlation and prediction of dense fluid transport coefficients. II. Simple molecular fluids. *Fluid Phase Equilib.* **1992**, *75*, 245.

(46) Dymond, J. H.; Awan M. A. Correlation of high-pressure diffusion and viscosity coefficients for n-alkanes. *Int. J. Thermophys.* **1989**, *10*, 941.

(47) Assael, M. J.; Dalaouti N. K.; Dymond J. H.; Feleki F. P. Correlation for Viscosity and Thermal Conductivity of Liquid Halogenated Ethane Refrigerants. *Int. J. Thermophys.* **2000**, *21*, 367.

(48) Chung, T. H.; Ajlan M.; L.L.Lee; Starling K. E. Generalized Multiparameter Correlation for Nonpolar and Polar Fluid Transport Properties. *Ind. Eng. Chem. Res.* **1988**, *27*, 671.

**Table 1.** Data sources for reference fluids

| Liquid | σ(η) | Temperature range / K | Pressure range / MPa |
|---|---|---|---|
| Propane[28] | ± 2% | 90-600 | 0.01-100 |
| Pentane[32] | ± 3% | 318-443 | 10-70 |
| Pentane[33] | ± 0.5% | 303.15-323.15 | 0.1-250 |
| Pentane[34] | ± 2% | 280-455 | 0.1-60 |
| Hexane[33]* | ± 0.5% | 303-348 | 0.1-250 |
| Heptane[35] | ± 2% | 298-348 | 0.1-170 |
| Heptane[36] | ± 0.5% | 303-348.15 | 0.1-250 |
| Toluene[22] | ± 0.5% | 298.15-373.15 | 0.1-200 |
| Toluene[37] | ± 0.5% | 303.15-323.15 | 0.1-80 |

σ(η) reported viscometer uncertainty, ΔT and Δp temperature and pressure intervals studied by the authors.* This fluid has been used first to verify the viscosity calibration procedure and after it has been included as reference fluid for viscosity determination of HFC-134a + polyether mixtures.

**Table 2.** Experimental (up to 60 MPa) or Tait-extrapolated densities, $\rho_{lab}$, and predicted densities using the Patel-Teja EoS with a quadratic mixing rule.

| | T/K | | | | | | | | | | | | | | |
|---|---|---|---|---|---|---|---|---|---|---|---|---|---|---|---|
| | 293.15 | | | 313.15 | | | 333.15 | | | 353.15 | | | 373.15 | | |
| p/MPa | $\rho_{lab}$ | $\rho_{EoS}$ | $\Delta\rho/\rho$ (%) | $\rho_{lab}$ | $\rho_{EoS}$ | $\Delta\rho/\rho$ (%) | $\rho_{lab}$ | $\rho_{EoS}$ | $\Delta\rho/\rho$ (%) | $\rho_{lab}$ | $\rho_{EoS}$ | $\Delta\rho/\rho$ (%) | $\rho_{lab}$ | $\rho_{EoS}$ | $\Delta\rho/\rho$ (%) |
| | **88% HFC-134a + 12% TriEGDME** | | | | | | | | | | | | | | |
| 10 | 1.2401 | 1.2525 | -1.0 | 1.1888 | 1.2000 | -0.9 | 1.134 | 1.1403 | -0.6 | 1.075 | 1.0721 | 0.3 | 1.0088 | 1.1010 | -9.1 |
| 20 | 1.2659 | 1.2831 | -1.4 | 1.2203 | 1.2396 | -1.6 | 1.1735 | 1.1921 | -1.6 | 1.125 | 1.1402 | -1.3 | 1.0744 | 1.1585 | -7.8 |
| 30 | 1.2877 | 1.3072 | -1.5 | 1.2462 | 1.2697 | -1.9 | 1.204 | 1.2295 | -2.1 | 1.1613 | 1.1865 | -2.2 | 1.1179 | 1.1992 | -7.3 |
| 40 | 1.3068 | 1.3270 | -1.5 | 1.268 | 1.2938 | -2.0 | 1.2294 | 1.2587 | -2.4 | 1.1906 | 1.2216 | -2.6 | 1.1514 | 1.2307 | -6.9 |
| 50 | 1.3238 | 1.3438 | -1.5 | 1.2874 | 1.3139 | -2.1 | 1.2511 | 1.2824 | -2.5 | 1.215 | 1.2496 | -2.8 | 1.1791 | 1.2563 | -6.5 |
| 60 | 1.3392 | 1.3583 | -1.4 | 1.3046 | 1.3309 | -2.0 | 1.2703 | 1.3024 | -2.5 | 1.2363 | 1.2728 | -3.0 | 1.2028 | 1.2777 | -6.2 |
| 70 | 1.3599 | 1.3709 | -0.8 | 1.3272 | 1.3457 | -1.4 | 1.2953 | 1.3195 | -1.9 | 1.2639 | 1.2925 | -2.3 | 1.2330 | 1.2960 | -5.1 |
| 80 | 1.3724 | 1.3821 | -0.7 | 1.3409 | 1.3587 | -1.3 | 1.3101 | 1.3345 | -1.9 | 1.2802 | 1.3095 | -2.3 | 1.2507 | 1.3120 | -4.9 |
| 90 | 1.3841 | 1.3921 | -0.6 | 1.3536 | 1.3703 | -1.2 | 1.3239 | 1.3477 | -1.8 | 1.2952 | 1.3245 | -2.3 | 1.2669 | 1.3261 | -4.7 |
| 100 | 1.3950 | 1.4012 | -0.4 | 1.3654 | 1.3806 | -1.1 | 1.3366 | 1.3594 | -1.7 | 1.3090 | 1.3377 | -2.2 | 1.2819 | 1.3387 | -4.4 |
| 110 | 1.4053 | 1.4094 | -0.3 | 1.3765 | 1.3900 | -1.0 | 1.3485 | 1.3700 | -1.6 | 1.3219 | 1.3496 | -2.1 | 1.2958 | 1.3500 | -4.2 |
| 120 | 1.4151 | 1.4169 | -0.1 | 1.3870 | 1.3985 | -0.8 | 1.3598 | 1.3796 | -1.5 | 1.3340 | 1.3604 | -2.0 | 1.3088 | 1.3602 | -3.9 |
| 130 | 1.4244 | 1.4237 | 0.0 | 1.3969 | 1.4063 | -0.7 | 1.3704 | 1.3884 | -1.3 | 1.3455 | 1.3701 | -1.8 | 1.3211 | 1.3696 | -3.7 |
| 140 | 1.4333 | 1.4301 | 0.2 | 1.4063 | 1.4134 | -0.5 | 1.3804 | 1.3964 | -1.2 | 1.3563 | 1.3791 | -1.7 | 1.3327 | 1.3781 | -3.4 |
| | **86% HFC-134a + 14% TEGDME** | | | | | | | | | | | | | | |
| 10 | 1.2417 | 1.2509 | -0.7 | 1.1920 | 1.2018 | -0.8 | 1.1391 | 1.1462 | -0.6 | 1.0824 | 1.0828 | 0.0 | 1.0193 | 1.1059 | -8.5 |
| 20 | 1.2662 | 1.2791 | -1.0 | 1.2219 | 1.2382 | -1.3 | 1.1759 | 1.1935 | -1.5 | 1.1288 | 1.1447 | -1.4 | 1.0798 | 1.1591 | -7.3 |
| 30 | 1.2869 | 1.3014 | -1.1 | 1.2462 | 1.2661 | -1.6 | 1.2048 | 1.2281 | -1.9 | 1.1633 | 1.1875 | -2.1 | 1.1209 | 1.1973 | -6.8 |
| 40 | 1.3051 | 1.3199 | -1.1 | 1.2671 | 1.2885 | -1.7 | 1.2290 | 1.2552 | -2.1 | 1.1910 | 1.2201 | -2.4 | 1.1529 | 1.2270 | -6.4 |

**Table 2.** *Continued*

| | T/K | | | | | | | | | | | | | | |
|---|---|---|---|---|---|---|---|---|---|---|---|---|---|---|---|
| | 293.15 | | | 313.15 | | | 333.15 | | | 353.15 | | | 373.15 | | |
| p/MPa | $\rho_{lab}$ | $\rho_{EoS}$ | $\Delta\rho/\rho$ (%) | $\rho_{lab}$ | $\rho_{EoS}$ | $\Delta\rho/\rho$ (%) | $\rho_{lab}$ | $\rho_{EoS}$ | $\Delta\rho/\rho$ (%) | $\rho_{lab}$ | $\rho_{EoS}$ | $\Delta\rho/\rho$ (%) | $\rho_{lab}$ | $\rho_{EoS}$ | $\Delta\rho/\rho$ (%) |
| | 86% HFC-134a + 14% TEGDME | | | | | | | | | | | | | | |
| 50 | 1.3214 | 1.3355 | -1.1 | 1.2854 | 1.3072 | -1.7 | 1.2499 | 1.2775 | -2.2 | 1.2145 | 1.2463 | -2.6 | 1.1792 | 1.2511 | -6.1 |
| 60 | 1.3360 | 1.3490 | -1.0 | 1.3020 | 1.3232 | -1.6 | 1.2683 | 1.2962 | -2.2 | 1.2349 | 1.2680 | -2.7 | 1.2022 | 1.2713 | -5.7 |
| 70 | 1.3558 | 1.3609 | -0.4 | 1.3238 | 1.3370 | -1.0 | 1.2924 | 1.3122 | -1.5 | 1.2615 | 1.2865 | -2.0 | 1.2311 | 1.2887 | -4.7 |
| 80 | 1.3677 | 1.3714 | -0.3 | 1.3369 | 1.3492 | -0.9 | 1.3068 | 1.3263 | -1.5 | 1.2772 | 1.3025 | -2.0 | 1.2482 | 1.3038 | -4.5 |
| 90 | 1.3789 | 1.3808 | -0.1 | 1.3492 | 1.3601 | -0.8 | 1.3201 | 1.3387 | -1.4 | 1.2916 | 1.3166 | -1.9 | 1.2638 | 1.3172 | -4.2 |
| 100 | 1.3894 | 1.3893 | 0.0 | 1.3606 | 1.3698 | -0.7 | 1.3324 | 1.3497 | -1.3 | 1.3050 | 1.3291 | -1.8 | 1.2782 | 1.3291 | -4.0 |
| 110 | 1.3993 | 1.3970 | 0.2 | 1.3713 | 1.3786 | -0.5 | 1.3440 | 1.3597 | -1.2 | 1.3175 | 1.3403 | -1.7 | 1.2916 | 1.3398 | -3.7 |
| 120 | 1.4086 | 1.4040 | 0.3 | 1.3814 | 1.3866 | -0.4 | 1.3549 | 1.3687 | -1.0 | 1.3292 | 1.3504 | -1.6 | 1.3042 | 1.3495 | -3.5 |
| 130 | 1.4175 | 1.4105 | 0.5 | 1.3910 | 1.3940 | -0.2 | 1.3652 | 1.3770 | -0.9 | 1.3402 | 1.3597 | -1.5 | 1.3160 | 1.3584 | -3.2 |
| 140 | 1.4260 | 1.4165 | 0.7 | 1.4001 | 1.4007 | 0.0 | 1.3750 | 1.3846 | -0.7 | 1.3507 | 1.3681 | -1.3 | 1.3272 | 1.3666 | -3.0 |

**Table 3.** Dynamic viscosity, η/mPa·s, for HFC-134a + polyether mixtures at different temperatures and pressures.

| | T/K | | | | |
|---|---|---|---|---|---|
| **88% HFC-134a + 12% TriEGDME** | | | | | |
| p /MPa | 293.15 | 313.15 | 333.15 | 353.15 | 373.15 |
| 10 | 0.404 | 0.330 | 0.256 | 0.203 | 0.160 |
| 20 | 0.445 | 0.362 | 0.288 | 0.233 | 0.196 |
| 30 | 0.486 | 0.395 | 0.320 | 0.263 | 0.228 |
| 40 | 0.529 | 0.429 | 0.352 | 0.291 | 0.257 |
| 50 | 0.573 | 0.465 | 0.384 | 0.319 | 0.284 |
| 60 | 0.619 | 0.502 | 0.415 | 0.347 | 0.309 |
| 70 | 0.664 | 0.539 | 0.446 | 0.373 | 0.329 |
| 80 | 0.712 | 0.579 | 0.477 | 0.401 | 0.354 |
| 90 | 0.762 | 0.620 | 0.508 | 0.428 | 0.376 |
| 100 | 0.813 | 0.663 | 0.540 | 0.454 | 0.398 |
| 110 | 0.868 | 0.707 | 0.575 | 0.479 | 0.416 |
| 120 | 0.922 | 0.754 | 0.606 | 0.505 | 0.437 |
| 130 | 0.980 | 0.802 | 0.637 | 0.530 | 0.457 |
| 140 | 1.039 | 0.852 | 0.669 | 0.556 | 0.478 |
| **86% HFC-134a + 14% TEGDME** | | | | | |
| 10 | 0.406 | 0.321 | 0.249 | 0.198 | 0.163 |
| 20 | 0.449 | 0.359 | 0.292 | 0.241 | 0.201 |
| 30 | 0.491 | 0.396 | 0.329 | 0.277 | 0.234 |
| 40 | 0.535 | 0.434 | 0.363 | 0.307 | 0.263 |
| 50 | 0.580 | 0.472 | 0.394 | 0.335 | 0.290 |
| 60 | 0.626 | 0.510 | 0.424 | 0.362 | 0.317 |
| 70 | 0.671 | 0.553 | 0.453 | 0.388 | 0.337 |
| 80 | 0.720 | 0.582 | 0.482 | 0.416 | 0.362 |
| 90 | 0.769 | 0.623 | 0.516 | 0.439 | 0.385 |
| 100 | 0.820 | 0.661 | 0.543 | 0.464 | 0.409 |
| 110 | 0.875 | 0.699 | 0.578 | 0.488 | 0.427 |
| 120 | 0.929 | 0.738 | 0.610 | 0.513 | 0.449 |
| 130 | 0.985 | 0.778 | 0.643 | 0.539 | 0.471 |
| 140 | 1.042 | 0.817 | 0.678 | 0.565 | 0.492 |

**Table 4**. Parameters and obtained deviations for Geller and Davis method for pure compounds

| | **HFC-134a** | | **TriEGDME** | | **TEGDME** | |
|---|---|---|---|---|---|---|
| p/MPa | B/ mPa$^{-1}$·s$^{-1}$ | $V_0$/cm$^3$·mol$^{-1}$ | B/ mPa$^{-1}$·s$^{-1}$ | $V_0$/cm$^3$·mol$^{-1}$ | B/ mPa$^{-1}$·s$^{-1}$ | $V_0$/cm$^3$·mol$^{-1}$ |
| 20 | 13.3312 | 62.4613 | 10.5285 | 172.3751 | 8.1106 | 211.6079 |
| 40 | 13.0378 | 62.1438 | 10.0856 | 171.1255 | 7.6823 | 210.0311 |
| 60 | 12.5540 | 61.5264 | 9.6075 | 169.8969 | 7.1924 | 208.4850 |
| 80 | 11.8765 | 60.7663 | 9.0784 | 168.7100 | 6.7532 | 207.0663 |
| 100 | 12.0781 | 60.7365 | 8.6763 | 167.6601 | 6.3425 | 205.7148 |
| **Deviations** | | | | | | |
| **AAD %** | 0.5 | | 2.1 | | 2.9 | |
| **DMAX %** | 1.0 | | 4.3 | | 6.3 | |
| **Bias %** | 0.01 | | -0.2 | | -0.4 | |

**Table 5.** Self-referencing model parameters and obtained deviations for pure compounds with $T_0$=293.15 K and $p_0$=20 MPa.

| Parameters | HFC-134a | TriEGDME | TEGDME |
|---|---|---|---|
| a | -0.3143 | 0.5199 | 0.6085 |
| b | -0.6440 | -0.4139 | -0.9432 |
| c | 0.4831 | 1.4749 | 1.7586 |
| d | -4.7946 | 10.5147 | 10.5393 |
| e | 47.8851 | -11.8065 | -11.7845 |
| f | 136.4407 | 191.1123 | 191.1260 |
| g | 71.5487 | -17.9536 | -17.9536 |
| h | 396.0128 | 446.2398 | 446.2398 |
| i | 1384.0740 | 1244.9188 | 1244.9188 |
| **Deviations** | | | |
| AAD % | 0.8 | 0.6 | 0.6 |
| DMAX % | 3.0 | 2.8 | 2.8 |
| Bias % | 0.1 | 0.0 | 0.0 |

**Table 6.** Parameter values and deviations obtained of free volume model for pure compounds.

| Pure Fluid | $\alpha$ /m$^5 \cdot$mol$^{-1} \cdot$s$^{-2}$ | B | $\ell$/Å |
|---|---|---|---|
| TriEGDME | 194.090 | 0.00341 | 0.390 |
| TEGDME | 255.546 | 0.00265 | 0.322 |
| HFC-134a | 23.970 | 0.01223 | 0.998 |
| **Deviations** | **AAD %** | **Bias %** | **DMAX %** |
| TriEGDME | 0.7 | -0.08 | 3.2 |
| TEGDME | 1.3 | -0.7 | 4.2 |
| HFC-134a | 0.8 | 0.0 | 2.1 |

**Table 7.** Parameter values and obtained deviations for pure compounds with the friction-theory + PR EoS

| Parameters | $a_0$/ μP·bar$^{-1}$ | $a_1$/ μP·bar$^{-1}$ | $a_2$/ μP·bar$^{-1}$ | $b_0$/ μP·bar$^{-1}$ | $b_1$/ μP·bar$^{-1}$ | $b_2$/ μP·bar$^{-1}$ | $c_2$/ μP·bar$^{-2}$ |
|---|---|---|---|---|---|---|---|
| TriEGDME | 12.1720 | -24.9801 | 4.5256 | 6.5219 | -18.8717 | 3.1219 | 5.1440 10$^{-5}$ |
| TEGDME | 36.4209 | -57.2566 | 9.0587 | 23.2056 | -40.2344 | 5.2509 | 8.8046 10$^{-5}$ |
| HFC-134a | 0.9572 | 5.0591 | -1.0432 | -0.7534 | 11.339 | -4.2165 | 1.1352 10$^{-5}$ |
| **Deviations** | AAD % | | | DMAX % | | Bias % | |
| TriEGDME | 1.2 | | | 4.1 | | 0.05 | |
| TEGDME | 1.8 | | | 4.5 | | 0.10 | |
| HFC-134a | 0.6 | | | 4.0 | | 0.06 | |

**FIGURE CAPTIONS**

**Figure 1.** Deviations between the experimental dynamic viscosity values of hexane reported by Oliveira and Wakeham[33], and the values determined using the three calibration methods: literature experimental data, (□), the hard-sphere method, (○) and the f-theory method, (△).

**Figure 2.** Viscosity values for HFC-134a + polyether mixtures in the temperature range of 293.15 K to 373.15 K and pressures between 10 MPa and 140 MPa using the three calibration methods: literature experimental data, (△), the hard-sphere method, (○) and the f-theory method, (◊). (a) Mixture containing TriEGDME and (b) mixture with TEGDME.

**Figure 3.** ηTp diagram for the HFC-134a + polyether (this work), and for pure HFC-134a[38] and pure TriEDGME and TEGDME[42].

**Figure 4.** Hard-Sphere parameters $V_0(T)$ versus temperature: (□) TEGDME[42], (■) TriEGDME[42], (▲) HFC-134a (eq. 13), (○) HFC-134a + TriEGDME and (◊) HFC-134a + TEGDME.

**Figure 5.** Deviations between the experimental values and the predicted data for the analysed models: (■) AAD, (□) DMAX, (■) Bias. (a) HFC-134a + TriEGDME and (b) HFC-134a + TEGDME.

**FIGURE 1**

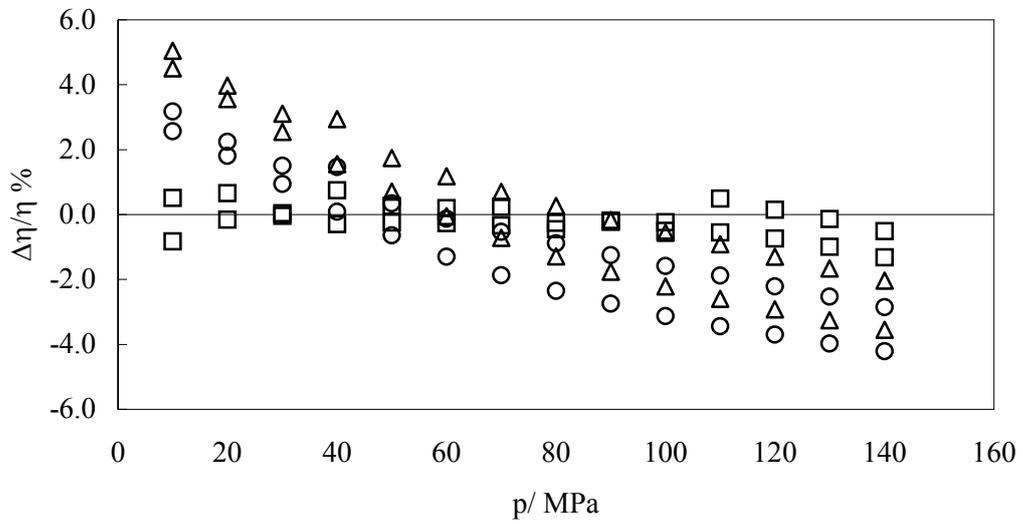

**FIGURE 2**

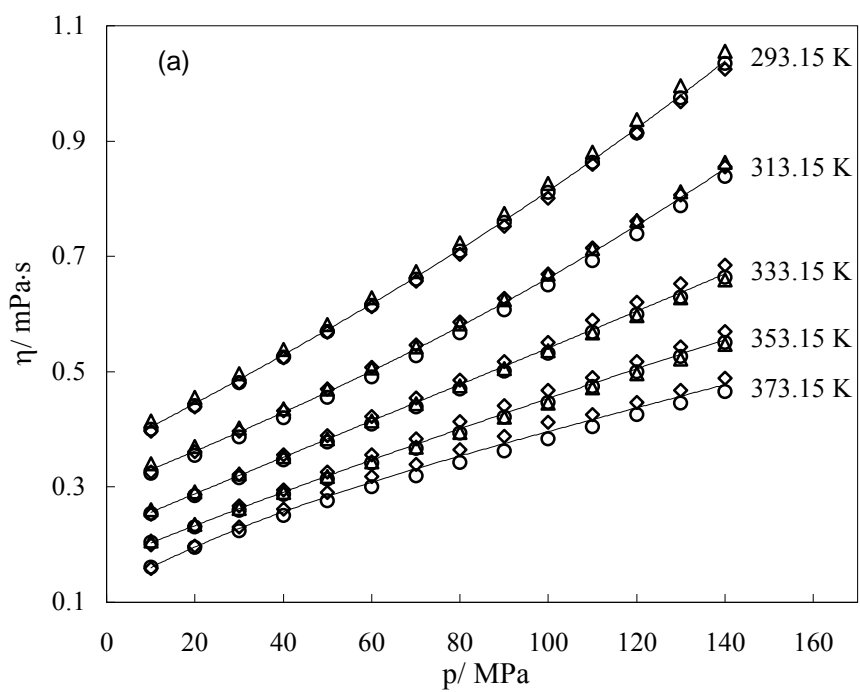

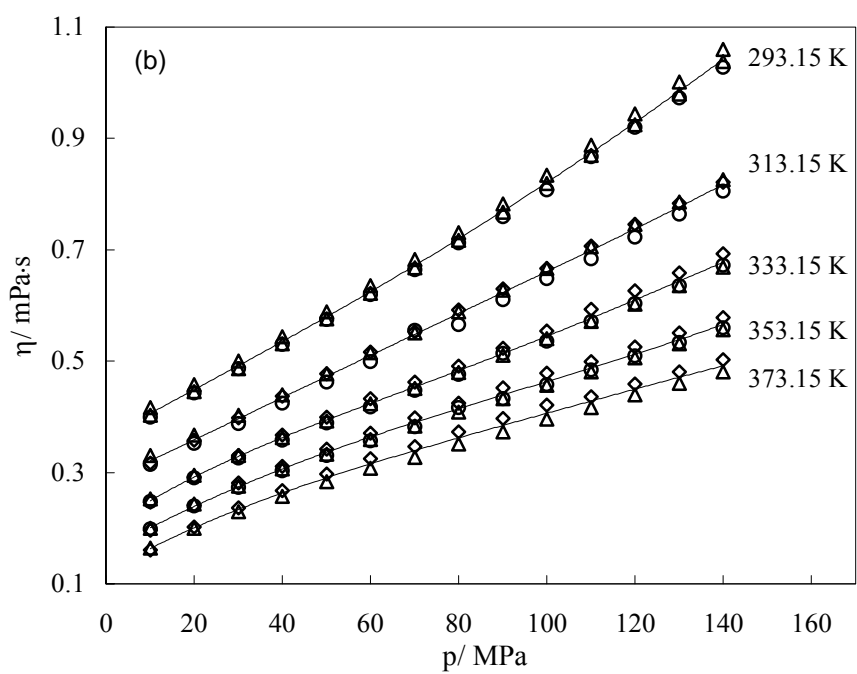

**FIGURE 3**

Pure TEGDME
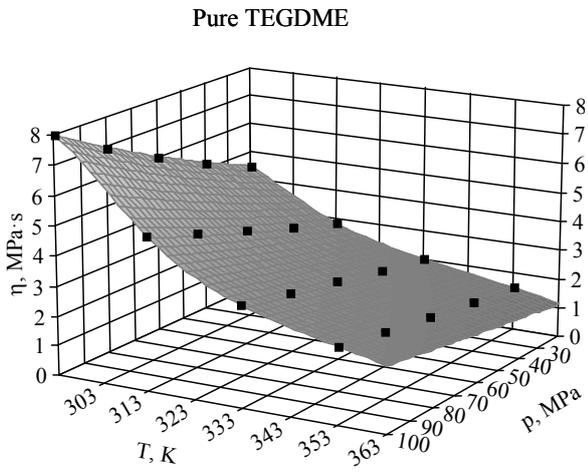

Pure TriEGDME
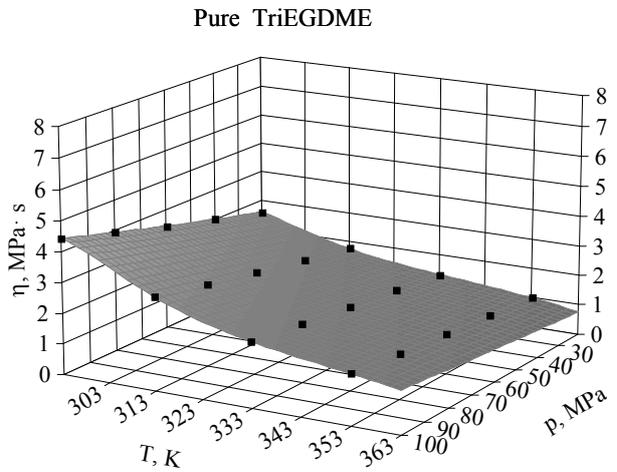

Pure HFC-134a
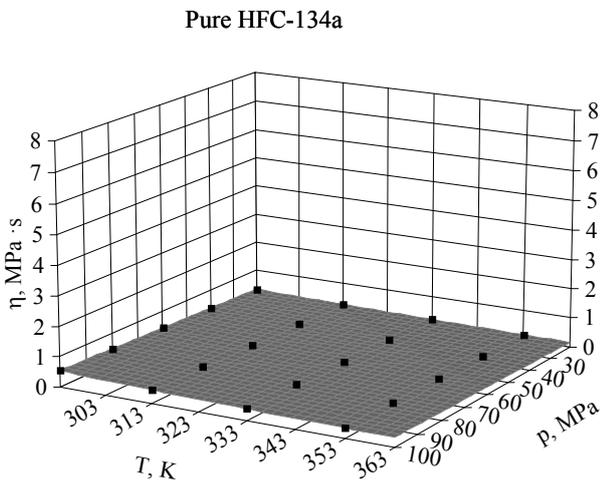

88 % HFC-134a + 12 % TriEGDME
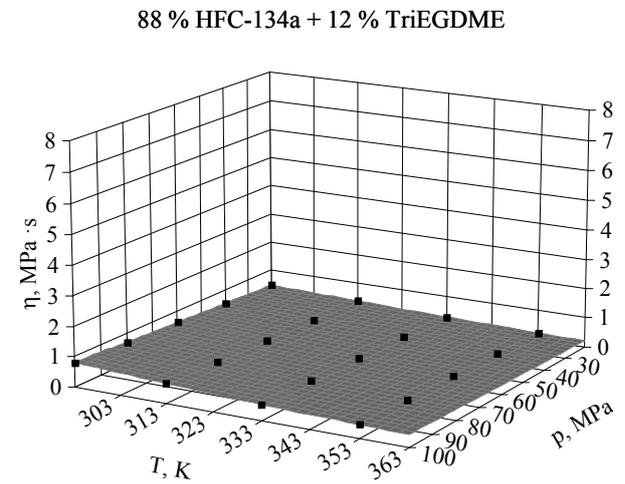

86 % HFC-134a + 14 % TEGDME
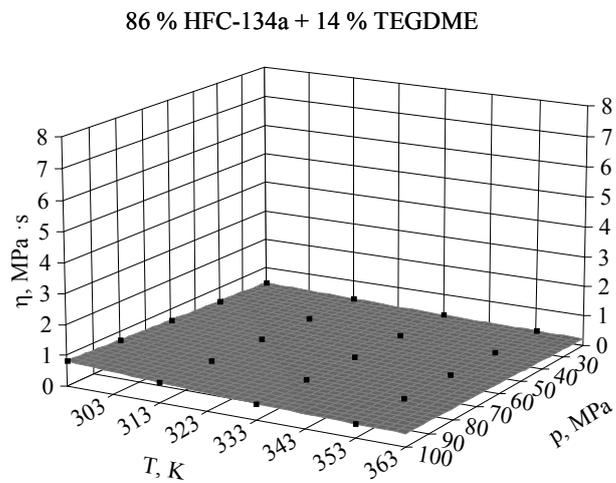

**FIGURE 4**

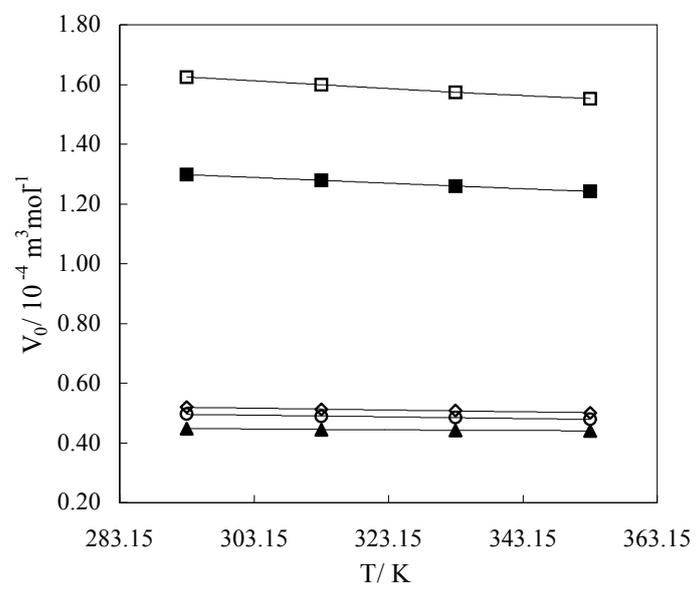

**FIGURE 5**

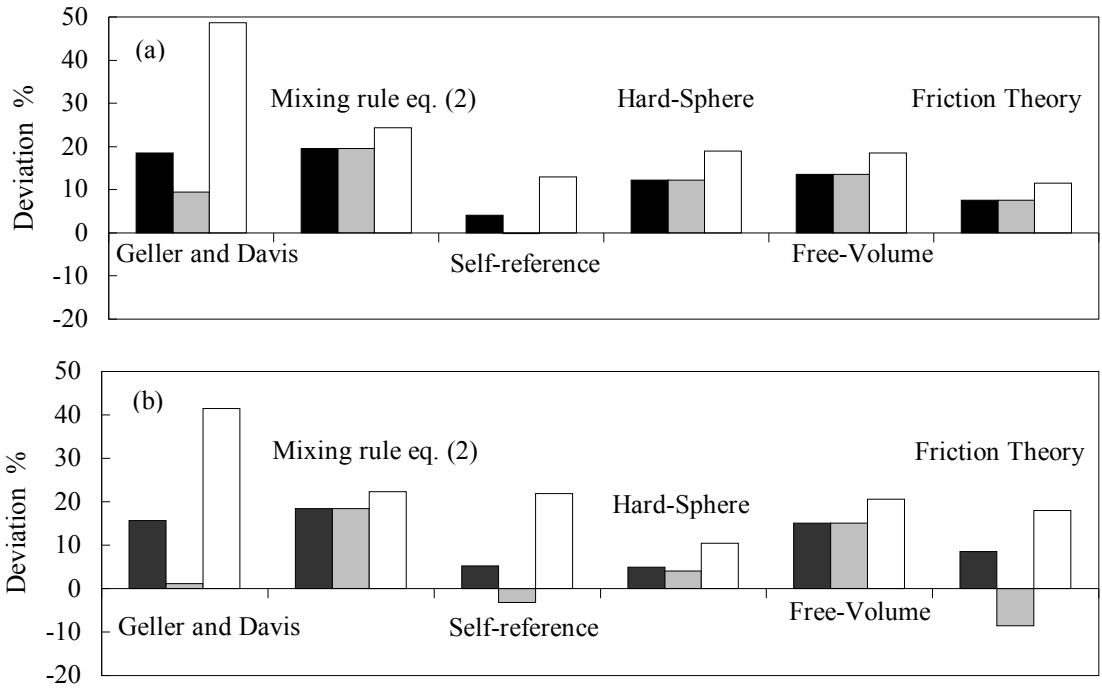